# Lateral recoil optical forces on nanoparticles near nonreciprocal surfaces

*Nayan K. Paul and J. Sebastian Gomez-Diaz*

Department of Electrical and Computer Engineering, University of California, Davis, One Shields Avenue, Kemper Hall 2039, Davis, CA 95616, USA

ABSTRACT: We investigate lateral recoil forces exerted on nanoparticles located near plasmonic platforms with in-plane nonreciprocal response. To this purpose, we first develop a comprehensive theoretical framework based on the Lorentz force within the Rayleigh approximation combined with nonreciprocal Green's functions and then derive approximate analytical expressions to model lateral recoil forces, demonstrating their explicit dependence on the dispersion relation of the system and unveiling the mechanisms that govern them. In particular, a dominant lateral recoil force component appears due to the momentum imbalance of nonreciprocal surface plasmons supported by the platform. This force can be orders of magnitude larger than other recoil force components, acts only along or against the direction of the external bias, and is quasi-independent of the direction, polarization, and wavelength of the incident plane wave. Lateral recoil forces are explored using drift-biased graphene metasurfaces, a platform that is also proposed to sort nanoparticles as a function of their size. Nonreciprocal plasmonic systems may enable new venues to trap, bind, and manipulate nanoparticles and to alleviate some of the challenges of conventional optical tweezers.

## I. INTRODUCTION

Lateral recoil optical forces over plasmonic structures have gathered large interest in recent years, as they enable the trapping [1-6] and manipulation [7-13] of nanoscale objects and can find important applications in bioengineering and chemistry [14-21]. Their underlying mechanism relies on the spin-orbit interaction of light [22]: upon adequate illumination, dipolar particles located near a plasmonic structure scatters quasi-circularly polarized (CP) light that couples to the surface in the form of directional surface plasmon polaritons (SPPs) [7-11]. To compensate for the momentum imbalance, the particle experiences a lateral recoil optical force acting in the direction opposite to the one of the excited plasmons [7-11] with a strength proportional to the momentum of the excited plasmons. The properties of the incident laser beam are critical in this process. For instance, the beam wavelength must be tuned to the system plasmon resonance to generate significant forces [11], whereas its polarization state influences the polarization acquired by the particle and determines the properties of the scattered light [10]. When these scattered evanescent waves are linearly polarized (LP), the excited SPPs propagate symmetrically within the surface and lateral recoil forces vanish. Potential approaches to achieve directional SPP excitation with LP light rely on the use of magneto-dielectric particles, such as Janus and Huygens dipoles [23], chiral objects [24] and high-index particles [25-28] that exploit combined effects of electric and magnetic dipole moments [29-31]. Unfortunately, these complex particles are not commonly found in chemical and biological applications.

An alternative route to overcome these challenges is using nonreciprocal plasmonic surfaces, for instance using magneto-optic materials biased with a magnetic field [31-36]. When the applied external bias is perpendicular to the surface, the polarization symmetry of the supported plasmons is broken. There, a LP nanoparticle can experience lateral recoil forces [31] because the scattered light undergoes a polarization conversion that excites directional SPPs. The strength of this recoil force is similar to the one found in reciprocal plasmonic systems [7] whereas its direction depends on the incident angle of the laser beam. When the applied external magnetic bias is parallel to the surface, the supported modes exhibit a broken symmetry in both amplitude and polarization [37-48]. Such structures have been shown to exert fluctuation-induced recoil forces on polarized atoms located nearby [49-52], in which the main emission channel is associated to the excitation of unidirectional SPPs supported at the material interface. Even though this platform allows to control the strength and direction of the induced forces with the external bias and the atomic transition frequency, fluctuation-induced forces are usually weak and thus uncapable of manipulating nanoparticles in practice.

Here, we investigate lateral optical recoil forces exerted on nanoparticles located near nonreciprocal interfaces illuminated with a plane wave. We focus on plasmonic platforms with an in-plane nonreciprocity that appears by applying an external bias parallel to the surface, and that manifests itself by a broken symmetry of the amplitude and polarization profile of the supported plasmons, as happens in the case of drift-biased graphene [37-42] and thin metals [43] or externally biased magneto-optic materials [33,44,46,47]. These general class of linear, homogeneous, anisotropic, plasmonic, and nonreciprocal metasurfaces possess negligible out-of-plane electric field responses and can be characterized using an effective nonlocal conductivity tensor with a broken symmetry in the momentum **k**-space, i.e., $\bar{\sigma}(\omega, \mathbf{k}) \neq \bar{\sigma}(\omega, -\mathbf{k})$ [37-40]. Alternatively, a nonlocal electrical permittivity tensor can also be employed. The electromagnetic behavior of these structures is distinct to the one obtained in electrical metasurfaces characterized with a local, fully populated conductivity tensor which may support hybrid surface plasmons with nonreciprocal responses in phase [31,48] but are unable to provide the full range of recoil forces discussed here.

For the sake of simplicity, and without loss of generality, we focus here on plasmonic metasurfaces characterized by a conductivity tensor $\bar{\boldsymbol{\sigma}}(\omega, \mathbf{k}) = \begin{bmatrix} \sigma_{xx}(\omega, \mathbf{k}) & 0 \\ 0 & \sigma_{yy}(\omega, \mathbf{k}) \end{bmatrix}$ and that are externally-biased within the plane. As a case-study, we consider a platform composed of a graphene layer longitudinally biased to generate drifting electrons with velocity $\boldsymbol{v}_d = v_d \hat{\boldsymbol{y}}$ along the surface, as illustrated in Figure 1. The graphene sheet is transferred onto a dielectric substrate with relative permittivity $\varepsilon_2$ and a nanoparticle is suspended in free space above the surface. This broadband nonreciprocal plasmonic platform has been recently experimentally demonstrated [41,42] and supports plasmons with unique features in the infrared band of the spectrum [37-40]. The isofrequency contour (IFC) of the supported modes is illustrated in Figure 1b. The applied bias breaks the rotational symmetry of SPPs in the momentum space and originates a nonreciprocal plasmonic response. Specifically, modes propagating against drifting electrons possess larger momentum than the ones in the opposite direction. Figure 1c shows that the system broken symmetry in the momentum space increases with the velocity of drifting electrons. This platform permits to engineer nonreciprocal responses over a broadband frequency ranging from terahertz to mid-infrared frequencies, as shown in Figure 1d. In case that the bias is applied along any other direction within the plane, this response can be captured by applying an adequate coordinate rotation. Even though we use drift-biased graphene as a platform to derive our theoretical framework, it should be stressed that our approach is general in the sense that it can readily be applied to describe any plasmonic surface with an in-plane nonreciprocal response.

In the following, we develop a theoretical formalism based on the Lorentz's force combined with the dyadic Green's functions of nonreciprocal surfaces to calculate the optical forces induced on nanoparticles located nearby. Then, we derive analytical expressions for all recoil force components by solving the dyadic Green's functions using the integration through the imaginary axis technique [49] combined with the residue theorem [50]. Our approach reveals that the dispersion relation of a plasmonic system suffices to analytically calculate all recoil force components, shedding light into the underlying mechanisms that enable them and facilitating the easy and accurate design of platforms capable of manipulating nanoparticles. Additionally, we show that the momentum imbalance of nonreciprocal SPPs leads to a dominant lateral recoil force component that acts along/against the applied bias and is mostly independent of the properties of the incoming laser beam. We study these forces over the drift-biased graphene platform described in Figure 1, exploring the strength and direction of recoil forces versus the velocity of electrons flowing on graphene, the polarization, frequency, and direction of the incoming light, and the particle position over the platform. We also investigate the capability of this platform to sort nanoparticles as a function of their size. The exciting properties of lateral recoil forces over nonreciprocal plasmonic surfaces are very promising for trapping, binding, and manipulating nanoparticles using low-power laser beams.

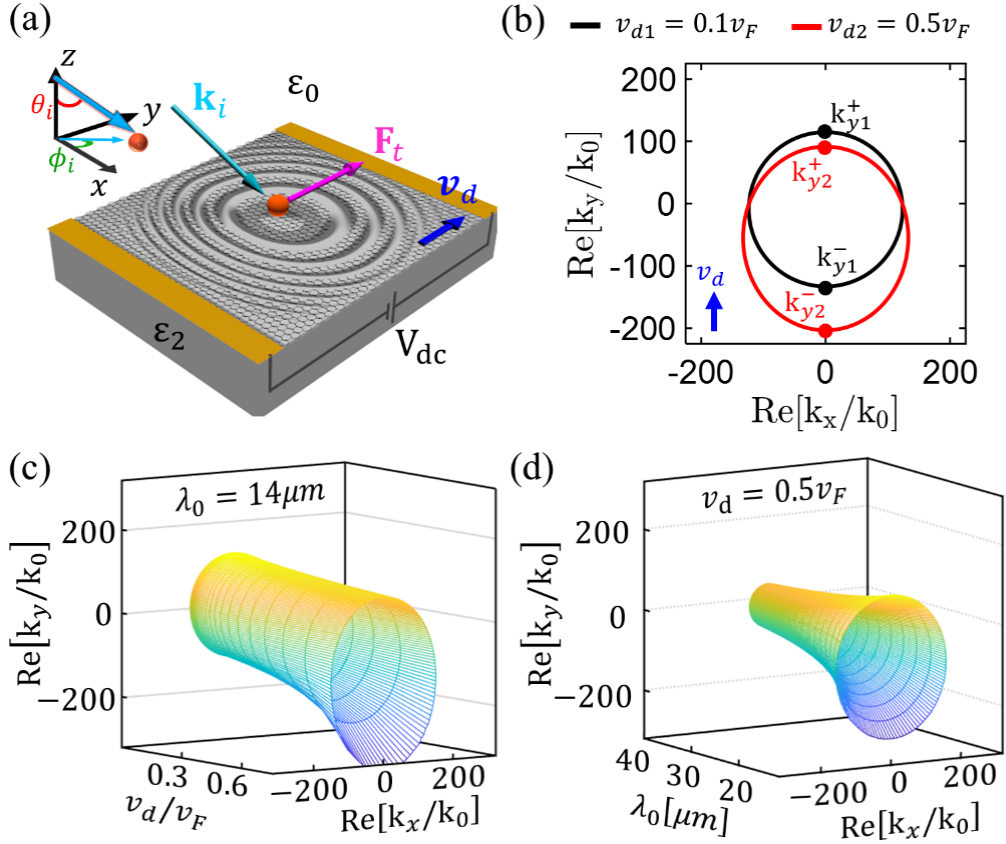

**Figure 1.** Lateral recoil optical forces acting on a nanoparticle located over a nonreciprocal plasmonic platform. (a) Schematic of the configuration. Lateral optical forces (magenta) are exerted on a Rayleigh particle (orange) located over a drift-biased graphene transferred over hexagonal boron nitride upon illumination with a plane wave (cyan). (b) Isofrequency contour of the states supported by graphene at $\lambda_0 = 14$ μm for two velocities of drifting electrons. $k_y^+$ and $k_y^-$ denote the supported states along and against the applied drift. (c)-(d) Momentum of the states supported by the platform versus the velocity of drifting electrons flowing along the graphene sheet and wavelength, respectively. Graphene's chemical potential and relaxation time are set to $\mu_c = 0.1 eV$ and $\tau = 0.3 ps$.

**II. LATERAL RECOIL FORCES NEAR NONRECIPROCAL SURFACES**

In this section, we derive a theoretical framework to compute lateral recoil optical forces acting on dipolar Rayleigh particles (radius $a < \lambda_0/20$, where $\lambda_0$ is wavelength) located near nonreciprocal plasmonic metasurfaces when illuminated by a laser beam. The system nonreciprocity is obtained by applying an external momentum bias parallel to its surface that can be designed in practice by using magnetic bias [32-34], moving metasurfaces [51], or drift-current bias [37-43]. Our proposed theory leads to compact expressions of the force components that depend on the Green's functions of the plasmonic surface, and is general in the sense that no assumptions regarding the type of plasmonic metasurface, surrounding media, operation frequency, or material of the Rayleigh particle are made.

The time-averaged total optical force exerted on a spherical dipolar Rayleigh particle in an arbitrary system can be computed as $\mathbf{F} = \frac{1}{2}\text{Re}\{\mathbf{p}^* \cdot \nabla \mathbf{E}^{\text{loc}}(\mathbf{r}_0, \mathbf{r}_0)\}$ [52]. Here, 'Re' is the real part of a complex number; '∗' is the complex conjugate; $\nabla$ is the vector gradient; $\mathbf{E}^{\text{loc}}$ is the total electric field at the particle position $\mathbf{r}_0$; and $\mathbf{p}$ is the effective dipole moment acquired by the particle. When the particle is located above a plasmonic medium and is illuminated with light, the total electric field $\mathbf{E}^{loc}$ at the particle position is composed of two terms [7-11]: the incident electric field together with the one reflected from the surface, and the evanescent field scattered by the particle that might couple to the surface in the form of SPPs. Calculating the dyadic Green's functions of the system, as described in Appendix A, permits to obtain the scattered fields and to determine the effective dipole moment $\mathbf{p}$ acquired by the particle (Appendix B).

Let us now consider that a plasmonic surface is subjected to an external momentum bias parallel to its surface, and that, as a result, it exhibits a nonreciprocal response. The total optical forces induced on a nanoparticle located above such surface can be decomposed as $\mathbf{F} = \mathbf{F}^0 + \mathbf{F}^{rec}$, where $\mathbf{F}^0$ is the conservative force component and $\mathbf{F}^{rec}$ gathers all nonconservative recoil forces. The conservative force arises from the gradient of the electric field intensity of the standing wave formed above the surface due to the superposition of incident field and the reflected one; whereas nonconservative forces depend on the properties of the evanescent fields scattered by the particle that couple to the surface in the form of SPPs. In the near field of the surface, conservative forces are usually orders of magnitude smaller than recoil forces [7-10], and their contribution to the net lateral optical forces can be safely neglected. Therefore, we focus here on the study of nonconservative lateral recoil forces. For completeness, we provide expressions to calculate the conservative optical forces component in Appendix C.

Nonconservative recoil optical forces can be computed from the gradient of the scattered electric field as

$$\mathbf{F}^{rec} = \frac{1}{2}\mathrm{Re}\{\mathbf{p}^* \cdot \nabla \mathbf{E}^s\}. \tag{1}$$

Here, $\mathbf{E}^s = \omega^2 \mu_0 \overline{\mathbf{G}}^S \cdot \mathbf{p}$ is the electric field of the excited SPPs at the dipole position $\mathbf{r}_0 = x_0\hat{x} + y_0\hat{y} + z_0\hat{z}$, $\omega = 2\pi f$ is the radial frequency, f is the operation frequency, $\mu_0$ is the free space permeability, and $\overline{\mathbf{G}}^S$ is the scattered dyadic Green's functions of the plasmonic system. Taking into account the nonzero derivatives of $\overline{\mathbf{G}}^S$ (see Appendix A), the identities $\frac{d}{dx}G_{xy}^s = \frac{d}{dx}G_{yx}^s$, $\frac{d}{dx}G_{xz}^s = -\frac{d}{dx}G_{zx}^s$, $\frac{d}{dy}G_{yz}^s = -\frac{d}{dy}G_{zy}^s$, $\frac{d}{dz}G_{yz}^s = -\frac{d}{dz}G_{zy}^s$ at the dipole position, and applying the approach detailed in [7,10], the lateral recoil force components from Eq. (1) can be expressed as

$$F_x^{rec} = \frac{k_0^2}{\varepsilon_0}\left[\text{Re}\{p_x^* p_y\}\text{Re}\left\{\frac{d}{dx}G_{xy}^s\right\}\right] - \frac{k_0^2}{\varepsilon_0}\text{Im}\{p_x^* p_z\}\text{Im}\left\{\frac{d}{dx}G_{xz}^s\right\}, \quad (2a)$$

$$F_y^{rec} = \frac{k_0^2}{2\varepsilon_0}\sum_{n=x,y,z}|p_n|^2\text{Re}\left\{\frac{d}{dy}G_{nn}^s\right\} - \frac{k_0^2}{\varepsilon_0}\text{Im}\left[p_y^* p_z\right]\text{Im}\left\{\frac{d}{dy}G_{yz}^s\right\}. \quad (2b)$$

Here, 'Im' is the imaginary part of a complex number, $k_0 = \omega\sqrt{\mu_0\varepsilon_0}$ is the free space wavenumber, $\omega$ is the radial frequency of the incoming wave, and $\varepsilon_0$ is the electrical permittivity of free space. Eq. (2) can be further decomposed as $\mathbf{F}_t^{rec} = \mathbf{F}_t^{nr} + \mathbf{F}_t^s = \hat{x}(F_x^{nr-p} + F_x^s) + \hat{y}(F_y^{nr-a} + F_y^s)$ that allows to classify recoil forces into two main groups: one associated to the broken symmetry of the nonreciprocal system $\mathbf{F}_t^{nr} = \hat{x}F_x^{nr-p} + \hat{y}F_y^{nr-a}$ and another associated to spin-orbit effects, $\mathbf{F}_t^s = \hat{x}F_x^s + \hat{y}F_y^s$ [7-11]. The superscripts 'a' and 'p' in $\mathbf{F}_t^{nr}$ refer to the broken symmetry in amplitude and polarization of the SPPs supported by the surface, respectively.

The lateral recoil forces $\mathbf{F}_t^{nr}$ appearing due to the broken reciprocity of the system read

$$F_x^{nr-p} = \frac{k_0^2}{\varepsilon_0}\left[\text{Re}\{p_x^* p_y\}\text{Re}\left\{\frac{d}{dx}G_{xy}^s\right\}\right], \quad (3a)$$

$$F_y^{nr-a} = \frac{k_0^2}{2\varepsilon_0}\left[|p_x|^2\text{Re}\left\{\frac{d}{dy}G_{xx}^s\right\} + |p_y|^2\text{Re}\left\{\frac{d}{dy}G_{yy}^s\right\} + |p_z|^2\text{Re}\left\{\frac{d}{dy}G_{zz}^s\right\}\right]. \quad (3b)$$

Taking into account the power radiated by the $n = \{x, y, z\}$ component of the dipole, $P_{rad}^n = \frac{c_0 k_0^4}{12\pi\varepsilon_0}|p_n|^2$ [7,10], Eq. (3) can be simplified to

$$F_x^{nr-p} = \frac{6\pi}{c_0 k_0^2} P_{rad}^{xy} \chi_{xy} \text{Re}\left\{\frac{d}{dx} G_{xy}^s\right\}, \tag{4a}$$

$$F_y^{nr-a} = \frac{6\pi}{c_0 k_0^2} \sum_{n=x,y,z} P_{rad}^n \text{Re}\left\{\frac{d}{dy} G_{nn}^s\right\}, \tag{4b}$$

where $P_{rad}^{xy} = P_{rad}^x + P_{rad}^y$ is the power radiated by the $xy$-component of the dipole in free space; and $\chi_{xy} = 2\frac{\text{Re}\{p_x^* p_y\}}{|p_x|^2 + |p_y|^2}$ is the dipole's in-plane polarization factor [31]. Eq. (4) shows that these recoil forces mainly depend on the extent of the nonreciprocal response measured through the real part of the spatial derivative of the Green's function at the dipole position, which is strictly zero in reciprocal systems [7-11]. The excitation of SPPs with different wavenumber (polarization) profiles within the plane gives rise to a recoil force $F_y^{nr-a}$ ($F_x^{nr-p}$) directed along (orthogonal) to the bias axis. $F_x^{nr-p}$ is associated to the asymmetric polarization conversion of evanescent waves and is zero when the particle acquires a dipole moment oriented along or orthogonal to the bias direction and maximum when the dipole is linearly polarized at an angle $45^0$ with respect to the bias axis, a state that favors the polarization conversion process of the scattered fields [31].

The remaining lateral recoil forces arise from the spin-orbit effects, and can be expressed as

$$\mathbf{F}_t^s = -\hat{x}\frac{k_0^2}{\varepsilon_0}\text{Im}\{p_x^* p_z\}\text{Im}\left\{\frac{d}{dx}G_{xz}^s\right\} - \hat{y}\frac{k_0^2}{\varepsilon_0}\text{Im}[p_y^* p_z]\text{Im}\left\{\frac{d}{dy}G_{yz}^s\right\}. \tag{5}$$

Taking into account the power radiated by the dipole, Eq. (5) can be simplified to

$$\mathbf{F}_t^s = \frac{6\pi}{c_0 k_0^2}\left[\hat{x}P_{rad}^{xz}\eta_y\text{Im}\left\{\frac{d}{dx}G_{xz}^s\right\} + \hat{y}P_{rad}^{yz}\eta_x\text{Im}\left\{\frac{d}{dy}G_{yz}^s\right\}\right], \tag{6}$$

where $P_{rad}^{xz} = P_{rad}^{x} + P_{rad}^{z}$ and $P_{rad}^{yz} = P_{rad}^{y} + P_{rad}^{z}$ are the power radiated by the $xz$ and $yz$-components of the dipole in free space, respectively; and $\eta_{y(x)} = -2\frac{\text{Im}\{p_{x(y)}^* p_z\}}{|p_{x(y)}|^2 + |p_z|^2}$ is the dipole polarization helicity [7,10]. It is evident from Eq. (6) that $\mathbf{F}_t^s$ is maximum when the dipole acquires a quasi-circular polarization state (i.e., $\eta \approx \pm 1$) and vanishes when it is linearly polarized (i.e., $\eta \to 0$) [7-10]. Lateral recoil forces arising from spin-orbit effects are well known in the literature and have been extensively investigated in reciprocal plasmonic systems [7-11].

This approach can easily be extended for the case in which the external bias is applied along the $\hat{q}$-axis in-plane to the surface, leading to the following general expressions for lateral recoil forces

$$F_m^{nr-p} = \frac{6\pi}{c_0 k_0^2} P_{rad}^{mq} \chi_{mq} \text{Re}\left\{\frac{d}{dm} G_{mq}^s\right\}, \tag{7a}$$

$$F_q^{nr-a} = \frac{6\pi}{c_0 k_0^2} \sum_{n=x,y,z} P_{rad}^n \text{Re}\left\{\frac{d}{dq} G_{nn}^s\right\}, \tag{7b}$$

$$F_q^s = \frac{6\pi}{c_0 k_0^2} P_{rad}^{qz} \eta_m \text{Im}\left\{\frac{d}{dq} G_{qz}^s\right\}, \tag{7c}$$

where $\hat{m}$ is orthogonal to $\hat{q}$ within the surface plane. Even though the force expressions shown in Eq. (7) are in a compact form, calculating the derivatives of the Green's functions associated to a nonreciprocal plasmonic surface is usually a challenging task that requires dedicated numerical routines as well as advanced integration techniques in the complex plane. This restricts our intuitive understanding of recoil optical forces and how they are associated to the modes supported by the plasmonic surface.

# III. ANALYTICAL MODEL OF LATERAL RECOIL FORCES NEAR NONRECIPROCAL SURFACES

This section derives approximate and compact analytical expressions for lateral recoil forces acting on nanoparticles near nonreciprocal surfaces, establishing a fundamental link between the force response and the dispersion relation of the plasmonic system. To this purpose, we first solve the Green's functions derivatives of nonreciprocal platforms analytically by combining the imaginary axis integration technique [49] with the residue theorem [50]. Then, we employ those solutions in Eq. (7) to develop analytical expressions for recoil optical forces.

The approach to calculate the Green's functions of a nonreciprocal surface at the source position is described in Appendix A. We begin our analytical treatment by transforming these standard expressions [37] into polar coordinates $(k'_\rho, k'_\phi)$ using the identities $k_x = k'_\rho \cos k'_\phi$ and $k_y = k'_\rho \sin k'_\phi$, yielding

$$\bar{\mathbf{G}}^s(\mathbf{r}_0, \mathbf{r}_0) = \int_0^{2\pi} \bar{\mathbf{N}}(k'_\phi) dk'_\phi, \tag{8a}$$

$$\bar{\mathbf{N}}(k'_\phi) = \int_0^\infty \frac{\bar{\mathbf{X}}_s(k'_\rho, k'_\phi)}{M(k'_\rho, k'_\phi) D(k'_\rho, k'_\phi)} e^{i2k'_z z_0} dk'_\rho, \tag{8b}$$

where $\mathbf{k} = k'_\rho \hat{\rho} + k'_\phi \hat{\varphi} + k'_z \hat{z}$ is the wavevector in polar coordinates; $\bar{\mathbf{X}}_s(k'_\rho, k'_\phi)$ is a tensor that includes the reflection and cross-coupling of propagative and evanescent waves; $D(k'_\rho, k'_\phi)$ represents the dispersion relation of the system and determines the response of the supported SPPs; and $M(k'_\rho, k'_\phi)$ is associated to the medium surrounding the structure. In reciprocal systems, Eq. (8) exhibits a symmetrical behavior in both physical and momentum spaces, whereas such symmetry is broken in case of nonreciprocal platforms [37-40].

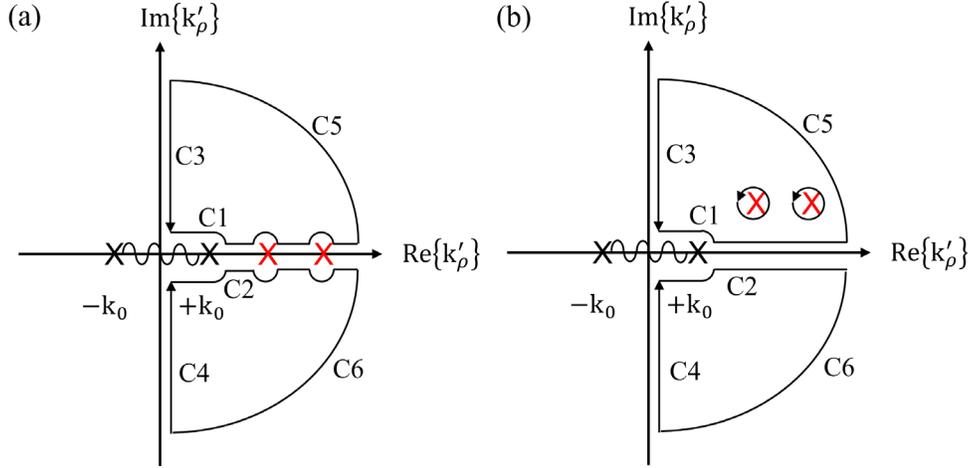

**Figure 2**. Proposed integration path in the complex plane for Eq. (8b) using the integration along the imaginary axis technique [49], considering (a) lossless and (b) lossy plasmonic platforms. Dominant poles, associated to SPPs, are represented with a red X.

To derive analytical expressions for the lateral recoil forces, we solve the integral along $k'_\rho$ shown in Eq. (8b) using the imaginary axis integration technique [49]. To this purpose, we assume that the system is low-loss and operated in the nonretarded regime (i.e., $k_\rho \gg k_0$, where $k_\rho$ is the wavenumber of the supported states). Figure 2 shows the integration strategy in the complex plane for a fixed $k'_\phi$, based on using six different integration paths denoted C1 to C6. First, we note that the Jordan's lemma is satisfied [50] and therefore the integrals around paths C5 and C6 when $k'_\rho \to \infty$ are strictly zero. Second, it can be shown [49] that the integrals around paths C1 and C2 are identical but with opposite sign, thus cancelling each other. Third, it should be noted that the dynamic part of the Green's functions tensor is determined by the surface modes supported by the platform [49] that appear in the form of poles in Figure 2 (red X). Assuming a lossless platform, the pole response is a real quantity that can be obtained analytically through the residue theorem [50]. These poles will dominate the response of lateral recoil forces in the platform. And forth, the integral around the paths C3 and C4 leads to a non-zero purely imaginary quantity associated to the quasi-static response of the Green's functions. Even though this integral is not analytical along the imaginary axis of $k'_\rho$, it is well-behaved and can be quickly integrated using numerical routines [53]. It should also be noted the integration contour has been deformed to avoid the brunch-cut from $-k_0$ to $+k_0$. In case of reciprocal platforms, the additional integration over the azimuthal angle $k'_\phi$ required in Eq. (8a) is simply a $2\pi$ constant. In case of nonreciprocal systems, such integration becomes angle dependent.

Applying this approach, the lateral recoil forces due to the system nonreciprocity can be approximated as

$$F_y^{nr-a} = \frac{\pi}{16\varepsilon_r c_0 k_0^4} \int_0^{2\pi} \frac{k_\rho^6 \sin^3 k'_\phi \cos^2 k'_\phi (A^x + B^{yz})}{(3\cos k'_\phi + 3\sin k'_\phi - 2\sqrt{2})} e^{-2z_0\sqrt{k_\rho^2 - k_0^2}} dk'_\phi, \quad (9a)$$

$$F_x^{nr-p} = \frac{3\pi P_{rad}^{yx} \chi_{yx}}{16\varepsilon_r c_0 k_0^4} \int_0^{2\pi} k_\rho^4 \cos k'_\phi \sin(2k'_\phi) e^{-2z_0\sqrt{k_\rho^2 - k_0^2}} dk'_\phi, \quad (9b)$$

where $\varepsilon_r = \frac{1+\varepsilon_2}{2}$ is the average of the relative permittivity of the media above ($\varepsilon_1 = 1$) and below ($\varepsilon_2$) the metasurface; the $k_\rho - k'_\phi$ relationship is given by the dispersion relation of the system $D(k'_\rho, k'_\phi)$; and the terms $A^x = \frac{1}{2} P_{rad}^x \cos^2 k'_\phi (\cos^2 k'_\phi + \sin k'_\phi)$ and $B^{yz} = (\sin k'_\phi + \cos k'_\phi)(P_{rad}^y \sin^2 k'_\phi + P_{rad}^z)$ accounts for the power scattered by the nanoparticle. Eq. (9) reveals that lateral recoil forces strongly depend on the dispersion relation of the modes supported by the platform. Even though these expressions are relatively complex, they exhibit a smooth variation versus $k'_\phi$ (see Appendix D) and can easily be integrated using standard numerical routines [53]. Intuitively, $F_y^{nr-a}$ dominates over $F_x^{nr-p}$ because it depends on the plasmon wavenumber $k_\rho$ with a power of six, whereas $F_x^{nr-p}$ shows a $\propto k_\rho^4$ dependence. $F_x^{nr-p}$ may still be relevant in case of magneto-optical substrates biased with a perpendicular magnetic field in which $F_y^{nr-a}$ vanishes [31].

Exploiting the dominant response along/against the external bias direction as described in Appendix D, the $F_y^{nr-a}$ component can be analytically approximated as

$$F_y^{nr-a} \approx \frac{3\pi P_{rad}^{yz}}{8c_0\varepsilon_r}\left[\left(\frac{k_y^-}{k_0}\right)^4 e^{-2z_0\sqrt{(k_y^-)^2-k_0^2}} - \left(\frac{k_y^+}{k_0}\right)^4 e^{-2z_0\sqrt{(k_y^+)^2-k_0^2}}\right], \quad (10)$$

where $k_y^-$ and $k_y^+$ are the wavenumber of the SPPs supported against/along the bias, as shown in Figure 1b. Eq. (10) shows that $F_y^{nr-a}$ depends on the fourth power of $k_y^\pm$, gets attenuated as $z_0$ increases, and is quasi-independent of the polarization state acquired by the particle. The only dependence of this force component with the direction, polarization, and wavelength of the incoming laser appears through the amount of scattered power, $P_{rad}^{yz}$. Eq. (10) reveals that this force is significantly enhanced as the asymmetry in the **k**-space along the bias direction increases and that it is zero in case of reciprocal media (i.e., $k_y^+ = k_y^-$). This equation also shows that nonreciprocity leads to a two-state system governed by the interplay between the distance $z_0$ and the momentum of the supported modes. Specifically, the force acts along the bias direction when the particle is close to the surface and excites confined SPPs with wavenumber $k_y^-$ that propagate against the bias. In this situation, the positive term of Eq. (10) dominates because $z_0 \to 0$ and $k_y^- > k_y^+$ (see Figure 1b). On the contrary, the force acts against the bias direction when the particle is located relatively far away from the surface. There, the high-**k** components of the scattered evanescent waves are filtered out by the free-space –modelled through the exponential terms in Eq. (10)– and cannot efficiently excite confined $k_y^-$ modes whereas they can still couple to the less confined $k_y^+$ states. The threshold distance $z_t$ at which the direction of $F_y^{nr-a}$ reverses can be approximated as

$$z_t \approx 2\frac{\ln(k_y^-)-\ln(k_y^+)}{k_y^- - k_y^+}, \quad (11)$$

which only depends on the plasmonic modes against/along the bias. It is important so establish the limits of our analytical model. Specifically, Eq. (10) holds when the polarization acquired by the particle along the direction orthogonal to the external bias is not dominant, and thus the scattered power fulfills $P_{rad}^x \leq P_{rad}^y$ and $P_{rad}^x \leq P_{rad}^z$ (see Appendix D). Such conditions are met in most practical scenarios as will be discussed in the following section. An exception appears when the nanoparticle is illuminated from the normal direction of the platform with light polarized in the direction orthogonal to the bias. There, Eq. (10) underestimates the strength of the forces as it does not account for the power of the fields scattered along that direction.

Following a similar procedure, described in Appendix D, spin-orbit lateral recoil forces can be obtained as

$$F_q^s = \frac{P_{rad}^{qz} \eta_m}{8\varepsilon_r c_0 k_0^4} \int_0^{2\pi} k_\rho^4 \, e^{-2z_0\sqrt{k_\rho^2 - k_0^2}} dk'_\phi, \tag{12}$$

and simplified to

$$F_y^s \approx \frac{3\pi P_{rad}^{yz} \eta_x}{8 c_0 \varepsilon_r} \left[ \left(\frac{k_y^+}{k_0}\right)^4 e^{-2z_0\sqrt{(k_y^+)^2 - k_0^2}} + \left(\frac{k_y^-}{k_0}\right)^4 e^{-2z_0\sqrt{(k_y^-)^2 - k_0^2}} \right], \tag{13a}$$

$$F_x^s \approx \frac{6\pi P_{rad}^{xz} \eta_y}{8 c_0 \varepsilon_r} \left(\frac{k_x}{k_0}\right)^4 e^{-2z_0\sqrt{k_x^2 - k_0^2}}. \tag{13b}$$

Here, $k_x$ is the wavenumber of the supported SPPs in the orthogonal lateral direction of the external bias. We stress that the analytical expressions of spin-orbit recoil forces shown in Eq. (13) holds for any dipole moment acquired by the particle in the scattering process. Inspecting Eq. (10) and (13), it becomes apparent that both nonreciprocal and spin-orbit recoil forces depend on the momentum of the plasmons supported by the platform. Main differences arise due to the underlying mechanisms that enable them: spin-orbit recoil forces mostly depend on the helicity acquired by the dipole ($\eta$) whereas nonreciprocal recoil forces rely on the broken symmetry of the supported SPPs along and against the external bias.

Finally, we remark that the analytical formalism described in this section assumes that the plasmonic surface is operated in the nonretarded regime (i.e., $k_\rho \gg k_0$). This approximation implies that the accuracy of the predicted recoil force components increases as the SPPs supported by the platform are more confined, and it holds independently of the position of the particle above the surface. We have numerically verified that this approximation holds well for $k_\rho \gtrsim 10 k_0$.

## IV. APPLICATION: LATERAL OPTICAL FORCES NEAR DRIFT-BIASED GRAPHENE

In this section, we investigate lateral recoil forces acting on a gold Rayleigh particle located near the drift-biased graphene transferred on hexagonal boron nitride described in Figure 1. The goals are threefold: (i) validate the accuracy of our analytical expressions for recoil forces; (ii) understand the behavior of optical forces appearing on nonreciprocal surfaces; and (iii) assess the possibility of sorting nanoparticles as a function of their size.

Figure 3 shows the strength of lateral recoil forces versus the electrons' drift velocity $v_d$, the properties of the incoming plane wave in terms of the azimuthal ($\phi_i$) and elevation angles ($\theta_i$), polarization, and wavelength, as well as the particle position $z_0$ over the metasurface. The force is normalized with respect to the power radiated by the particle's acquired dipole moment when it is located in free space, $\mathrm{P}_{\mathrm{rad}}^0 = \frac{c_0 k_0^2}{12\pi}\left(|\mathrm{p}_x|^2 + |\mathrm{p}_y|^2 + |\mathrm{p}_z|^2\right)$ [53]. Figure 3a shows the force components versus $v_d$ assuming a transverse magnetic (TM)-polarized incident light aligned with the $\hat{\mathrm{x}}$-axis (see Figure 1a). As expected, $\mathrm{F}_y^{\mathrm{nr-a}}$ strength increases with the applied bias and

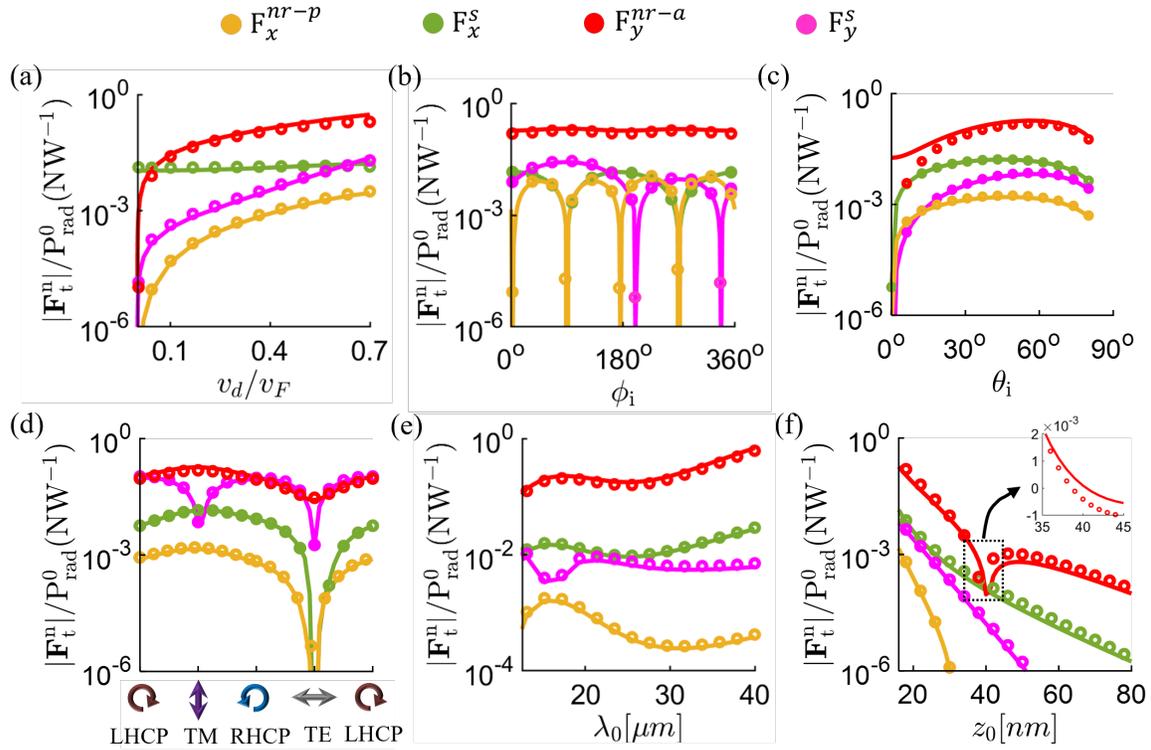

**Figure 3.** Normalized strength and direction of the lateral recoil force components acting on a gold nanoparticle with radius $a = 15$ nm located over the drift-biased graphene platform described in Figure 1. Results are calculated versus (a) the drift velocity of the flowing electrons; the properties of the incoming plane wave in terms of (b) azimuthal and (c) elevation angle of incidence; (d) polarization; (e) wavelength; and (f) the separation distance $z_0$ between the particle and the surface. Solid lines are computed numerically by solving Eq. (7) and markers analytically with Eqs. (10) and (13). Parameters that are not swept in a panel are kept to: $z_0 = a + 1$ [nm], $v_d = 0.5 v_F$, $\phi_i = 0^0$, $\theta_i = 60^0$, and light with TM-polarization.

outperforms all other force components by over an order of magnitude even with moderate drift velocities (i.e., $v_d \approx 0.2 v_F$ [54-56], where $v_F \approx 10^6$ m/s is graphene's Fermi velocity [57]), whereas $F_x^{nr-p}$ is negligible due to the weak polarization conversion in the system. Results confirm that the spin-orbit recoil force $F_x^s$ does not depend on the external bias whereas the orthogonal component $F_y^s$ increases with it, as described by Eq. (13). This is because larger bias enhances the momentum of the modes supported along the -*y* direction whereas it does not affect the modes supported on the orthogonal *x̂*-axis. This analysis holds even when the linearly polarized light comes from different azimuthal directions $\phi_i$, as shown in Figure 3b. Note that $F_x^{nr-p}$ vanishes in
17

the case of $\phi_i \approx \{0°, 90°, 180°, 270°\}$ because light polarization conversation does not take place in these cases [i.e., $\chi_{yx} \to 0$ in Eq. (7a)]. $F_y^{nr-a}$ remains dominant in all cases and exhibits a unidirectional response that does not depend on the beam direction. However, both the strength and direction of the spin-orbit force components depend on the laser-particle alignment due to their dependence on the polarization spin acquired by the nanoparticle. These forces vanish when the out-of-plane spin of the particle polarization is negligible [i.e., $\eta \to 0$ in Eq. (7c)]. Figure 3c explores the response of the recoil force components versus the elevation angle $\theta_i$. Maximum force strength is found over a relatively large angular range, roughly from 15° to 80° measured from the normal direction. When the particle is illuminated from the normal direction ($\theta_i \approx 0°$) with a TM-polarized laser beam, it acquires a dominant x-directed dipole moment. In that scenario, Eq. (10) underestimates the force $F_y^{nr-a}$ because it does not account for the power scattered in the direction orthogonal to the bias, $P_{rad}^x$. The polarization of the incoming light plays a critical role in this process, as it determines the spin acquired by the particle and the total power that it radiates. Figure 3d shows that $F_y^{nr-a}$ dominates when the incoming light is quasi-linearly polarized and reveals that, in case of quasi-CP light, spin orbit forces acquire a comparable strength due to the strong polarization spin ($\eta \sim \pm 1$) acquired by the particle, as stems from Eqs. (10) and (13). Therefore, the total recoil optical force is determined in these cases by the interplay between $F_y^{nr-a}$ and $F_y^s$, and the net optical forces over the system are not unidirectional anymore but may change with the laser angle of incidence. Again, we note that $F_x^{nr-p}$ vanishes for TE-polarized incoming light due to the absence of polarization conversion (i.e., $\chi_{yx} \to 0$) whereas spin-orbit forces vanish when the incoming light is linearly polarized (i.e., $\eta \to 0$). Figure 3e shows the recoil forces response versus operation wavelength. Again, $F_y^{nr-a}$ dominates even for laser beams oscillating over a wide frequency region in the infrared band. Such response arises because drift-biased graphene exhibits a broadband nonreciprocal behavior (see Figure 1d) in stark contrast with other reciprocal [37,38] and even nonreciprocal plasmonic systems [31,58,59] that require lasers tuned to their plasma frequency to provide significant recoil forces. Finally, Figure 3f studies the system response versus the particle position $z_0$ over the surface and confirms that $F_y^{nr-a}$ directs the nanoparticle along or against the applied bias with respect to the threshold position $z_t$ as described in Eq. (11). At exactly $z_0 = z_t$, the energy and momentum of SPPs flowing along and against the drift are equal thus yielding to $F_q^{nr-a} = 0$. It is important to emphasize that the analytical expressions shown in Eqs. (10) and (13) –markers in Figure 3– capture such complex responses of the recoil forces and agree very well with solutions obtained by numerically solving Eq. (7).

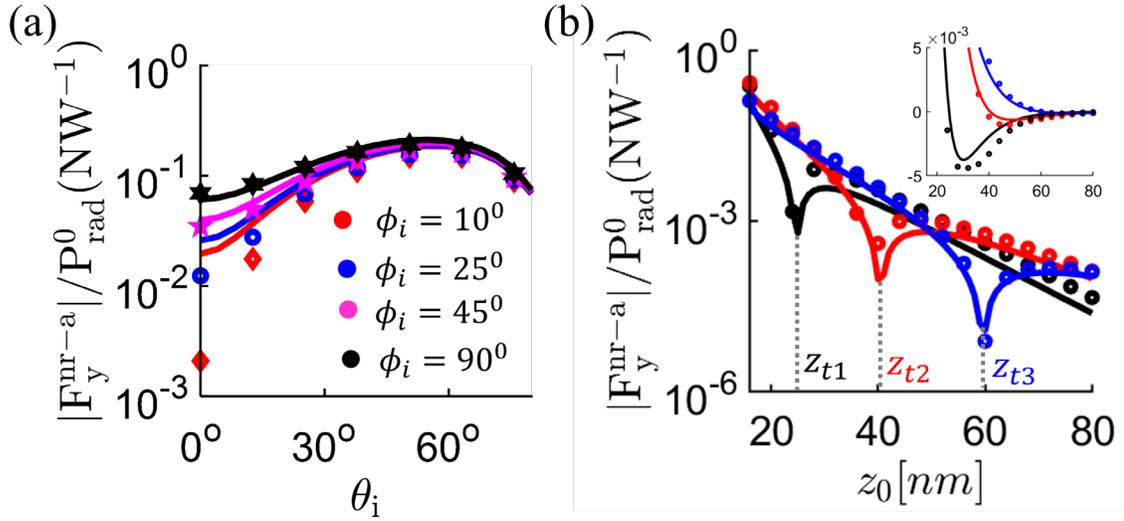

**Figure 4**: Normalized strength of the nonreciprocity-induced lateral recoil force $F_y^{nr-a}$ acting on a nanoparticle located on the drift-biased graphene platform described in Figure 1. Results are plotted versus (a) the elevation angle $\theta_i$ of the incoming light considering various azimuthal angles $\phi_i = 10°$ (red), $\phi_i = 25°$ (blue), $\phi_i = 45°$ (magenta) and $\phi_i = 90°$ (black); and (b) the particle position $z_0$ for different values of graphene's chemical potential: $\mu_c = 0.05 eV$ (black), $\mu_c = 0.1 eV$ (red) and $\mu_c = 0.15 eV$ (blue). Dashed lines denote the threshold distance $z_t$ in which the force direction reverses predicted by Eq. (11). Solid lines and markers correspond to the numerical [Eq. (7)] and analytical solutions [Eqs. (10) and (13)], respectively. Other parameters are as in Figure 1.

Figure 4 further explores the accuracy of the proposed analytical formulation to model nonreciprocity-induced lateral recoil forces. Specifically, Figure 4a shows the normalized strength of the $F_y^{nr-a}$ force component versus the elevation angle $\theta_i$ of the incident TM light for different azimuthal $\phi_i$ directions (see Figure 1). As described in Appendix D, Eq. (10) underestimates the recoil force response when the power radiated by the dipole along the direction orthogonal to the bias ($P_{rad}^x$) dominates over the other components ($P_{rad}^y$ and $P_{rad}^z$), a situation that appears in this platform when $\phi_i \lesssim 25^0$ and $\theta_i \lesssim 25^0$. Beyond these range of angles, $P_{rad}^x$ becomes comparable/weaker than $P_{rad}^y$ and $P_{rad}^z$ and thus its influence on the lateral recoil force becomes negligible, enabling the use of Eq. (10). In case that the polarization state of the incoming light is not TM (see Figure 3d), $P_{rad}^x$ is not dominant and thus Eq. (10) becomes again a very good approximation of the nonreciprocity-induced recoil force independently of the direction of the incident light. Figure 4b shows the normalized strength of $F_y^{nr-a}$ versus the particle position above the surface for three different values of graphene's chemical potential: $\mu_c = 0.05 eV$, $\mu_c = 0.1 eV$ and $\mu_c = 0.15 eV$. Results confirm that the direction of this optical force component along the external bias axis can be tuned in real time by adjusting graphene's gate-bias while barely affecting the overall force strength. It should also be stressed the good agreement obtain between numerical and analytical calculation (solid lines/markers) as well as the accuracy of Eq. (11) to predict the particle position in which the force direction changes its sign.

This nonreciprocal platform can readily be applied to sort nanoparticles as a function of their size. Figure 5a illustrates the normalized $z$-component of the electric field associated to the plasmons excited on the metasurface when it is biased with $\boldsymbol{v}_d = 0.5 v_F \hat{\boldsymbol{y}}$ and a gold nanoparticle with radius $a$ is located at $z_0 = a + 1[\text{nm}]$ over the system and is illuminated with a TM-plane wave. Top and bottom panels consider the case of a particle with radius $a = 20$ nm and $a = 50$ nm, respectively. In both cases, the particle scatters quasi-LP light (i.e., $\eta \to 0$) thus leading to negligible lateral spin-orbit forces $F_q^s$. In this scenario, the recoil force $F_y^{nr-a}$ arising from the broken symmetry of the platform dominates the optical forces, leading to the two-state system described in Eq. (10). When the particle is small (top panel), the confined $k_y^-$ plasmon propagating toward $-y$ is efficiently excited and the recoil force is exerted along the bias direction; however, when the particle size increases, $k_y^-$ states cannot be excited but the scattered light still couples to the less confined $k_y^+$ plasmons travelling toward the $+\hat{y}$-axis, which result in a recoil force acting against the bias direction. Figure 5b investigates the total forces acting on gold nanoparticles versus their size and the azimuthal direction of the incoming beam. As expected, the strength of the total optical force increases with the particle size as they scatter a larger amount of power that effectively couple to the system surface. Results show that particles with a radius larger than ~42 nm experience unidirectional forces against the drift, whereas those with a radius smaller than ∼37 nm are dragged along the applied bias. Such response is preserved independently of the direction of the incoming laser beam. The threshold between the two states of the system agrees well with the ∼38 nm predicted by Eq. (11). The particles with radii close to the threshold experience lateral forces that smoothly change direction with respect to $\phi_i$, a response that appears due to the interplay between spin-orbit and nonreciprocal recoil forces. Even though such changes can be very quick depending on the balance of the forces, for instance when the beam is coming close to $\phi_i = 150°$, the strength of the resulting total force in this scenario is insignificant. Finally, it should be noted that this sorting platform is dynamic in the sense that the particles radii threshold to direct them along/against the bias ($z_t$) can be manipulated in real time by changing graphene's Fermi level through a gate bias, as shown in Figure 4b.

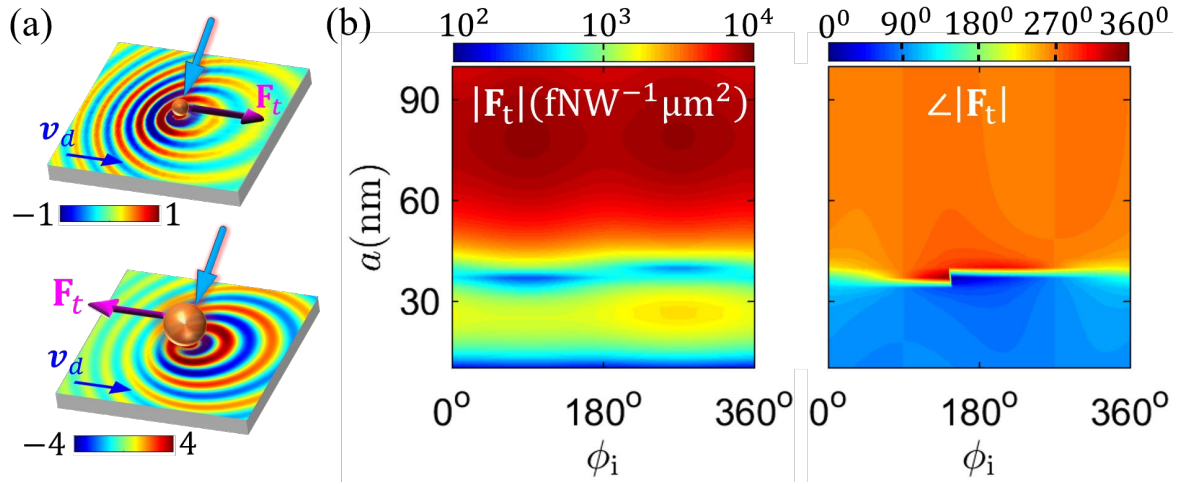

**Figure 5.** Lateral sorting of gold nanoparticles as a function of their radius $a$ using the platform described in Figure 1 with drift velocity $v_d = 0.5v_F$. The particles are located at a distance $z_0 = a + 1$ [nm] over a drift-biased graphene and illuminated with a TM-polarized plane wave at $\lambda_0 = 14\,\mu m$ coming from the elevation angle $\theta_i = 60^0$. (a) Normalized $z$-component of the surface plasmons excited on the platform for particles with radii 20 nm (top) and 50 nm (bottom). (b) Normalized total lateral force strength $|\mathbf{F}_t| = \sqrt{F_x^2 + F_y^2}$ (left) and direction $\angle|\mathbf{F}_t| = \tan^{-1}(F_y/F_x)$ (right) versus the particles' radii and azimuthal angle $\phi_i$ of the incident plane wave. Other parameters are as in Figure 1.

## V. CONCLUSIONS

In summary, we have explored lateral recoil forces acting on Rayleigh particles located over plasmonic platforms with in-plane nonreciprocal response. To this purpose, we have developed a comprehensive theoretical framework based on the Lorentz's force combined with the Green's functions that describe this type of systems. By applying the integration along the imaginary axis technique combined with the residue theorem, we solved these Green's functions analytically. The resulting approximate expressions establish a fundamental link between the lateral recoil forces acting on nanoparticles located near nonreciprocal plasmonic surfaces and the dispersion relation of the system. Additionally, in-plane nonreciprocity leads to a lateral recoil force component that only depends on the broken symmetry of the supported plasmons and the total optical power scattered by the particle, while being independent of any other property of the incoming light (wavelength, angle of incidence, polarization). Such force can be dominant over other recoil force components and creates a two-state system in which the particles are dragged along/against the external bias depending on their size and distance to the platform surface. Moving beyond, we envision that lateral recoil forces based on nonreciprocal platforms, including those composed of magneto-optical materials such as cobalt-silver alloy [31], topological gyrotropic materials [59], photonic topological insulators [58], or drift-biased 2D materials [37-42] and thin-metallic layers [43], will find numerous applications in physics, chemistry and bioengineering, with emphasis on alleviating some of the challenges of conventional optical tweezers in terms of photoheating [60], alignment, operation wavelength, and resolution to manipulate the lateral position of nanoparticles.

**FUNDING SOURCES**

This work was supported by the National Science Foundation with Grant No. ECCS-1808400 and a CAREER Grant No. ECCS-1749177.

**ACKNOWLEDGMENT**

The authors thank Dr. Alejandro Alvarez Melcon, professor of Telecommunication Engineering at the Technical University of Cartagena, Spain, for discussions regarding the analytical solutions of Green's functions using the residue theorem.

**APPENDIX A: DYADIC GREEN'S FUNCTIONS OF PLASMONIC NONRECIPROCAL SURFACES**

A useful model to study the electromagnetic response of linear and homogeneous plasmonic metasurfaces relies on using the Green's function formalism [52]. This approach has been studied in recent years for the case of nonreciprocal plasmonic systems, including graphene and metals [37-43]. In this appendix, we formulate the scattered dyadic Green's function of a nonreciprocal platform in the presence of an external in-plane momentum bias as a required step to calculate optical forces acting on nanoparticles within the Rayleigh approximation [52]. For the sake of simplicity, we consider that an external bias is applied on the system along $\hat{\mathbf{y}}$-axis of the reference coordinate system, as shown in Figure 1. In case that the bias is applied along a different direction, the formalism detailed below holds by applying a coordinate transformation.

| Reciprocal | Nonreciprocal |
|---|---|
| $\bar{\mathbf{G}}^s(\mathbf{r}_0, \mathbf{r}_0) = \begin{bmatrix} G_{xx}^s & 0 & 0 \\ 0 & G_{yy}^s & 0 \\ 0 & 0 & G_{zz}^s \end{bmatrix}$ | $\bar{\mathbf{G}}^s(\mathbf{r}_0, \mathbf{r}_0) = \begin{bmatrix} G_{xx}^s & 0 & 0 \\ 0 & G_{yy}^s & G_{yz}^s \\ 0 & G_{zy}^s & G_{zz}^s \end{bmatrix}$ |

**Table A1:** Zero and non-zero elements of the scattered dyadic Green's function tensor at the dipole position for reciprocal and nonreciprocal plasmonic surfaces. Nonreciprocity is obtained by applying an in-plane momentum bias along the $\hat{\mathbf{y}}$-axis of the plasmonic platform, as shown in Figure 1.

Let us consider an arbitrarily polarized electric dipole located at a position, $\mathbf{r}_0 = \{x_0, y_0, z_0\}$ above a nonreciprocal, anisotropic, structure. The scattered dyadic Green's functions of the system at the dipole position $\mathbf{r}_0$ can be expressed as [52]

$$\bar{\mathbf{G}}^s(\mathbf{r}_0, \mathbf{r}_0) = \frac{i}{8\pi^2} \iint_{-\infty}^{\infty} R_{ss}\bar{\mathbf{M}}_{ss} + R_{ps}\bar{\mathbf{M}}_{ps} + R_{sp}\bar{\mathbf{M}}_{sp} + R_{pp}\bar{\mathbf{M}}_{pp}]e^{i2k_z z_0} dk_x dk_y, \quad (A1)$$

where $\bar{\mathbf{R}}(k_x, k_x) = [R_{ss}(k_x, k_x) \; R_{sp}(k_x, k_x); R_{ps}(k_x, k_x) \; R_{pp}(k_x, k_x)]$ is the reflection coefficient tensor of the nonreciprocal structure [37-40] that can be calculated by solving Maxwell's boundary conditions [52]. Nonlocal effects are explicitly included and depend on the specific mechanism employed to bias the surface. $\bar{\mathbf{M}}$-matrices in Eq. (A1) are evaluated from the vector dyadic products of polarization unit vectors as described in [52]. In case of reciprocal structures, the nondiagonal elements of $\bar{\mathbf{G}}^s(\mathbf{r}_0, \mathbf{r}_0)$ are strictly zero due to the symmetric response of the guided modes supported the surface [7-11]. However, the applied bias breaks such symmetry in nonreciprocal systems, leading to a scattered dyadic Green's functions tensor that is not diagonal at the particle position. The zero and non-zero components of $\bar{\mathbf{G}}^s(\mathbf{r}_0, \mathbf{r}_0)$ in the case of reciprocal (unbiased) and nonreciprocal (external in-plane momentum bias applied along $\hat{\mathbf{y}}$-direction) structures are given in Table A1. Note that the tensor elements fulfill the following identity at the dipole position: $G^s_{yz}(\mathbf{r}_0,\mathbf{r}_0) = -G^s_{zy}(\mathbf{r}_0,\mathbf{r}_0)$.

| Reciprocal | Nonreciprocal |
|---|---|
| $\frac{d}{dx}\overline{\mathbf{G}}^s(\mathbf{r}_0,\mathbf{r}_0) = \begin{bmatrix} 0 & 0 & \frac{d}{dx}G^s_{xz} \\ 0 & 0 & 0 \\ \frac{d}{dx}G^s_{zx} & 0 & 0 \end{bmatrix}$ | $\frac{d}{dx}\overline{\mathbf{G}}^s(\mathbf{r}_0,\mathbf{r}_0) = \begin{bmatrix} 0 & \frac{d}{dx}G^s_{xy} & \frac{d}{dx}G^s_{xz} \\ \frac{d}{dx}G^s_{yx} & 0 & 0 \\ \frac{d}{dx}G^s_{zx} & 0 & 0 \end{bmatrix}$ |
| $\frac{d}{dy}\overline{\mathbf{G}}^s(\mathbf{r}_0,\mathbf{r}_0) = \begin{bmatrix} 0 & 0 & 0 \\ 0 & 0 & \frac{d}{dy}G^s_{yz} \\ 0 & \frac{d}{dy}G^s_{zy} & 0 \end{bmatrix}$ | $\frac{d}{dy}\overline{\mathbf{G}}^s(\mathbf{r}_0,\mathbf{r}_0) = \begin{bmatrix} \frac{d}{dy}G^s_{xx} & 0 & 0 \\ 0 & \frac{d}{dy}G^s_{yy} & \frac{d}{dy}G^s_{yz} \\ 0 & \frac{d}{dy}G^s_{zy} & \frac{d}{dy}G^s_{zz} \end{bmatrix}$ |

**Table A2:** Zero and non-zero elements of the $x$ and $y$-derivatives of the scattered dyadic Green's functions tensor at the dipole position for reciprocal and nonreciprocal plasmonic surfaces. Nonreciprocity is obtained by applying an in-plane momentum bias along the $\hat{\mathbf{y}}$-axis of the plasmonic platform, as shown in Figure 1.

To compute the lateral optical forces exerted on the electric point dipole, one needs to calculate the spatial derivatives of the Green's functions [7-11]. Such derivates yield

$$\frac{d}{dx}\overline{\mathbf{G}}^s(\mathbf{r}_0,\mathbf{r}_0) = -\frac{1}{8\pi^2}\iint_{-\infty}^{\infty} k_x\left[R_{ss}\overline{\mathbf{M}}_{ss} + R_{ps}\overline{\mathbf{M}}_{ps} + R_{sp}\overline{\mathbf{M}}_{sp} + R_{pp}\overline{\mathbf{M}}_{pp}\right]e^{i2k_z z_0}dk_x dk_y, \quad (A2a)$$

$$\frac{d}{dy}\overline{\mathbf{G}}^s(\mathbf{r}_0,\mathbf{r}_0) = -\frac{1}{8\pi^2}\iint_{-\infty}^{\infty} k_y\left[R_{ss}\overline{\mathbf{M}}_{ss} + R_{ps}\overline{\mathbf{M}}_{ps} + R_{sp}\overline{\mathbf{M}}_{sp} + R_{pp}\overline{\mathbf{M}}_{pp}\right]e^{i2k_z z_0}dk_x dk_y. \quad (A2b)$$

Most component of these tensors are strictly zero for reciprocal structures [7-11]. However, this situation changes in case of nonreciprocal systems. Table A2 compares the zero and non-zero derivatives of the Green's functions for reciprocal and nonreciprocal platforms. In this work, we solve Eqs. (A1) and (A2) using standard numerical routines that implement integration in the complex plane.

# APPENDIX B: EFFECIVE DIPOLE POLARIZABILITY OF NANOPARTICLES LOCATED NEAR NONRECIPROCAL SURFACES

In this section, we develop a formalism to compute the effective dipole polarizability induced on an isotropic, dipolar, spherical and Rayleigh particle located in free space over a linear and nonreciprocal plasmonic structure that is biased along the $\hat{y}$-axis. During the light-scattering process, the electric field scattered by the dipole at the particle position can be computed from the dyadic Green's functions as [52]

$$\mathbf{E}^s(\mathbf{r}_0, \mathbf{r}_0) = \omega^2 \mu_0 \begin{bmatrix} G_{xx}^s p_x \\ G_{yy}^s p_y + G_{yz}^s p_z \\ G_{zy}^s p_y + G_{zz}^s p_z \end{bmatrix}, \tag{B1}$$

where $\mathbf{p} = \hat{x} p_x + \hat{y} p_y + \hat{z} p_z$ is the effective dipole moment acquired by the particle calculated as [7-11]

$$\mathbf{p} = \alpha_0 \big[ \mathbf{E}^i(\mathbf{r}_0, \mathbf{r}_0) + \mathbf{E}^r(\mathbf{r}_0, \mathbf{r}_0) + \mathbf{E}^s(\mathbf{r}_0, \mathbf{r}_0) \big]. \tag{B2}$$

Here, $\alpha_0$ is the dynamic polarizability of the particle in free space [52] and $\mathbf{E}^i$ and $\mathbf{E}^r$ are the electric fields of the incident plane wave and the one reflected from the surface, respectively. Substituting $\mathbf{E}^s(\mathbf{r}_0, \mathbf{r}_0)$ into Eq. (B2) yields to the following set of equations:

$$(1 - \omega^2 \mu_0 \alpha_0 G_{xx}^s) p_x = \alpha_0 E_x^0, \tag{B3a}$$

$$(1 - \omega^2 \mu_0 \alpha_0 G_{yy}^s) p_y - \omega^2 \mu_0 \alpha_0 G_{yz}^s p_z = \alpha_0 E_y^0, \tag{B3b}$$

$$-\omega^2 \mu_0 \alpha_0 G_{zy}^s p_y + (1 - \omega^2 \mu_0 \alpha_0 G_{zz}^s) p_z = \alpha_0 E_z^0, \tag{B3c}$$

where $\mathbf{E}^0 = \mathbf{E}^i + \mathbf{E}^r$ is the superposition of the incident and reflected electric fields of the plane wave, and $\mu_0$ is the free space permeability. The elements of the dipole moment induced on the particle are then computed solving Eq. (B3), leading to

$$p_x = \frac{E_x^0 \alpha_0}{1-\omega^2 \mu_0 \alpha_0 G_{xx}^S}, \tag{B4a}$$

$$p_y = \frac{E_y^0 \alpha_0 (1-\omega^2 \mu_0 \alpha_0 G_{zz}^S) + E_z^0 \omega^2 \mu_0 \alpha_0^2 G_{yz}^S}{(1-\omega^2 \mu_0 \alpha_0 G_{yy}^S)(1-\omega^2 \mu_0 \alpha_0 G_{zz}^S) - \omega^4 \mu_0^2 \alpha_0^2 G_{yz}^S G_{zy}^S}, \tag{B4b}$$

$$p_z = \frac{E_y^0 \omega^2 \mu_0 \alpha_0^2 G_{zy}^S + E_z^0 \alpha_0 (1-\omega^2 \mu_0 \alpha_0 G_{yy}^S)}{(1-\omega^2 \mu_0 \alpha_0 G_{yy}^S)(1-\omega^2 \mu_0 \alpha_0 G_{zz}^S) - \omega^4 \mu_0^2 \alpha_0^2 G_{yz}^S G_{zy}^S}. \tag{B4c}$$

Following the identity $\mathbf{p} = \overline{\boldsymbol{\alpha}}^{\text{eff}} \cdot \mathbf{E}^0$ [7-11], the nonzero elements of the effective dipole polarizability tensor of the particle can be expressed as

$$\alpha_{xx}^{\text{eff}} = \frac{\alpha_0}{1-\omega^2 \mu_0 \alpha_0 G_{xx}^S}, \tag{B5a}$$

$$\alpha_{yy}^{\text{eff}} = \frac{\alpha_0 (1-\omega^2 \mu_0 \alpha_0 G_{zz}^S)}{(1-\omega^2 \mu_0 \alpha_0 G_{yy}^S)(1-\omega^2 \mu_0 \alpha_0 G_{zz}^S) - \omega^4 \mu_0^2 \alpha_0^2 G_{yz}^S G_{zy}^S}, \tag{B5b}$$

$$\alpha_{zz}^{\text{eff}} = \frac{\alpha_0 (1-\omega^2 \mu_0 \alpha_0 G_{yy}^S)}{(1-\omega^2 \mu_0 \alpha_0 G_{yy}^S)(1-\omega^2 \mu_0 \alpha_0 G_{zz}^S) - \omega^4 \mu_0^2 \alpha_0^2 G_{yz}^S G_{zy}^S}, \tag{B5c}$$

$$\alpha_{yz}^{\text{eff}} = \frac{\omega^2 \mu_0 \alpha_0^2 G_{yz}^S}{(1-\omega^2 \mu_0 \alpha_0 G_{yy}^S)(1-\omega^2 \mu_0 \alpha_0 G_{zz}^S) - \omega^4 \mu_0^2 \alpha_0^2 G_{yz}^S G_{zy}^S}, \tag{B5d}$$

$$\alpha_{zy}^{\text{eff}} = \frac{\omega^2 \mu_0 \alpha_0^2 G_{zy}^S}{(1-\omega^2 \mu_0 \alpha_0 G_{yy}^S)(1-\omega^2 \mu_0 \alpha_0 G_{zz}^S) - \omega^4 \mu_0^2 \alpha_0^2 G_{yz}^S G_{zy}^S}. \tag{B5e}$$

|  Reciprocal  |  Nonreciprocal  |
|---|---|
| $\overline{\boldsymbol{\alpha}}^{\text{eff}} = \begin{bmatrix} \alpha_{xx}^{\text{eff}} & 0 & 0 \\ 0 & \alpha_{yy}^{\text{eff}} & 0 \\ 0 & 0 & \alpha_{zz}^{\text{eff}} \end{bmatrix}$ | $\overline{\boldsymbol{\alpha}}^{\text{eff}} = \begin{bmatrix} \alpha_{xx}^{\text{eff}} & 0 & 0 \\ 0 & \alpha_{yy}^{\text{eff}} & \alpha_{yz}^{\text{eff}} \\ 0 & \alpha_{zy}^{eff} & \alpha_{zz}^{\text{eff}} \end{bmatrix}$ |

**Table B1**: Zero and non-zero elements of the effective dipole polarizability tensor for reciprocal and nonreciprocal plasmonic surfaces. Nonreciprocity is obtained by applying an in-plane momentum bias along the $\hat{y}$-axis of the plasmonic platform, as shown in Figure 1.

Note that the nondiagonal components $\alpha_{yz}^{\text{eff}}$ and $\alpha_{zy}^{\text{eff}}$ completely vanish for the case of reciprocal platforms [2-6]. Table B1 compares the zero and non-zero elements of the effective dipole polarizability tensors for reciprocal and nonreciprocal structures.

**APPENDIX C: CONSERVATIVE OPTICAL FORCES**

The lateral radiation pressure acting on the particle can be computed from the electric field of the incident and reflected plane waves as [52]

$$\mathbf{F}^0 = \frac{1}{2}\text{Re}\{\mathbf{p}^* \cdot \nabla \mathbf{E}^0\}. \tag{C1}$$

Following the procedure detailed in [7-11], the lateral ($q = \{x, y\}$) radiation pressure reads

$$F_q = \frac{1}{2}k_q^0 \left[ \text{Im}\{\alpha_{xx}^{\text{eff}}\}|E_x^0|^2 + \text{Im}\{\alpha_{yy}^{\text{eff}}\}|E_y^0|^2 + \text{Im}\{\alpha_{zz}^{\text{eff}}\}|E_z^0|^2 \right. \\ \left. + 2\text{Im}\{\alpha_{yz}^{\text{eff}}\}\text{Im}\{E_z^{0*}E_y^0\} \right], \tag{C2}$$

where $k_q^0$ is the lateral wavenumber of the illuminating light in free space, and the external bias has been applied along the $\hat{y}$-axis. Note that for the case of a reciprocal platform, the off-diagonal components of the particle polarizability are strictly zero (i.e., $\alpha_{yz}^{\text{eff}} = \alpha_{zy}^{\text{eff}} = 0$) and Eq. (C2) simplifies to the common one found over reciprocal surfaces [7-11].

# APPENDIX D: ANALYTICAL APPROXIMATION OF OPTICAL FORCES NEAR NONRECIPROCAL SURFACES

This appendix derives analytical expressions for the optical force components described in Eq. (7) by solving the Green's functions derivatives associated to nonreciprocal surfaces using the integration along the imaginary axis technique [49] combined with the residue theorem [50].

**D1. Lateral recoil forces due to the broken symmetry of the system in amplitude**

Let us consider first the lateral recoil forces that appear due to the broken symmetry in amplitude of the supported SPPs and described using Eq. (7b). Assuming that the external bias is applied along the $\hat{y}$-axis, this force component can be expressed as

$$F_y^{nr-a} = \frac{6\pi}{c_0 k_0^2} \left[ P_{rad}^x \text{Re}\left\{\frac{d}{dy} G_{xx}^s\right\} + P_{rad}^y \text{Re}\left\{\frac{d}{dy} G_{yy}^s\right\} + P_{rad}^z \text{Re}\left\{\frac{d}{dy} G_{zz}^s\right\} \right]. \tag{D1}$$

Following the integration path described in Figure 2, the integral along $k_\rho'$ of the Green's functions spatial derivatives at the particle position can be analytically computed by applying the residue theorem. In our notation, $k_\rho$ is the longitudinal component of the supported surface mode supported at $k_\phi'$. The properties of these modes can be obtained by solving the dispersion relation of the system given by $D(k_\rho', k_\phi') = 0$. This approach leads to analytical expressions for the residues that are somewhat lengthy and difficult to work with. To further simplify them and gain physical insight into the problem, we assume that the platform operates in the nonretarded regime ($k_\rho \gg k_0$ [7,11]). This permits to relate the effective conductivity along $x$ and $y$ directions with the wavenumber of the modes supported therein as $\sigma_{xx} \approx i\omega\varepsilon_0 \frac{2}{k_x}$ and $\sigma_{yy} \approx i\omega\varepsilon_0 \frac{2}{k_y}$. The resulting expressions are

$$N_{\frac{d}{dy}G_{xx}^s}(k_\phi') \approx -i \frac{k_\rho^6 \sin^3 k_\phi' \cos^4 k_\phi' \left(\cos^2 k_\phi' + \sin k_\phi'\right)}{192\pi\varepsilon_r k_0^2 \left(3\cos k_\phi' + 3\sin k_\phi' - 2\sqrt{2}\right)} e^{-2z_0 \sqrt{k_\rho^2 - k_0^2}}, \tag{D2a}$$

$$N_{\frac{d}{dy}G_{yy}^s}(k'_\phi) \approx -i \frac{k_\rho^6 \sin^5 k'_\phi \cos^2 k'_\phi (\sin k'_\phi + \cos k'_\phi)}{96\pi\varepsilon_r k_0^2 (3\cos k'_\phi + 3\sin k'_\phi - 2\sqrt{2})} e^{-2z_0\sqrt{k_\rho^2 - k_0^2}}, \qquad (D2b)$$

$$N_{\frac{d}{dy}G_{zz}^s}(k'_\phi) \approx -i \frac{k_\rho^6 \sin^3 k'_\phi \cos^2 k'_\phi (\sin k'_\phi + \cos k'_\phi)}{96\pi\varepsilon_r k_0^2 (3\cos k'_\phi + 3\sin k'_\phi - 2\sqrt{2})} e^{-2z_0\sqrt{k_\rho^2 - k_0^2}}. \qquad (D2c)$$

Next, the spatial derivatives of the Green's functions are computed at the particle position by performing polar integrals along $k'_\phi$:

$$\text{Re}\left\{\frac{d}{dy}G_{xx}^s\right\} = \int_0^{2\pi} \frac{k_\rho^6 \sin^3 k'_\phi \cos^4 k'_\phi (\cos^2 k'_\phi + \sin k'_\phi)}{192\varepsilon_r k_0^2 (3\cos k'_\phi + 3\sin k'_\phi - 2\sqrt{2})} e^{-2z_0\sqrt{k_\rho^2 - k_0^2}} dk'_\phi, \qquad (D3a)$$

$$\text{Re}\left\{\frac{d}{dy}G_{yy}^s\right\} = \int_0^{2\pi} \frac{k_\rho^6 \sin^5 k'_\phi \cos^2 k'_\phi (\sin k'_\phi + \cos k'_\phi)}{96\varepsilon_r k_0^2 (3\cos k'_\phi + 3\sin k'_\phi - 2\sqrt{2})} e^{-2z_0\sqrt{k_\rho^2 - k_0^2}} dk'_\phi, \qquad (D3b)$$

$$\text{Re}\left\{\frac{d}{dy}G_{zz}^s\right\} = \int_0^{2\pi} \frac{k_\rho^6 \sin^3 k'_\phi \cos^2 k'_\phi (\sin k'_\phi + \cos k'_\phi)}{96\varepsilon_r k_0^2 (3\cos k'_\phi + 3\sin k'_\phi - 2\sqrt{2})} e^{-2z_0\sqrt{k_\rho^2 - k_0^2}} dk'_\phi, \qquad (D3c)$$

where again the relationship between $k_\rho$ and $k'_\phi$ is implicit and given by the dispersion relation of the system. Note that these expressions are strictly zero in the case of reciprocal devices, due to the polar symmetry of the platform response. Combining Eqs. (D1) and (D3) permits to obtain the expression of this recoil force component as shown in Eq. (9a).

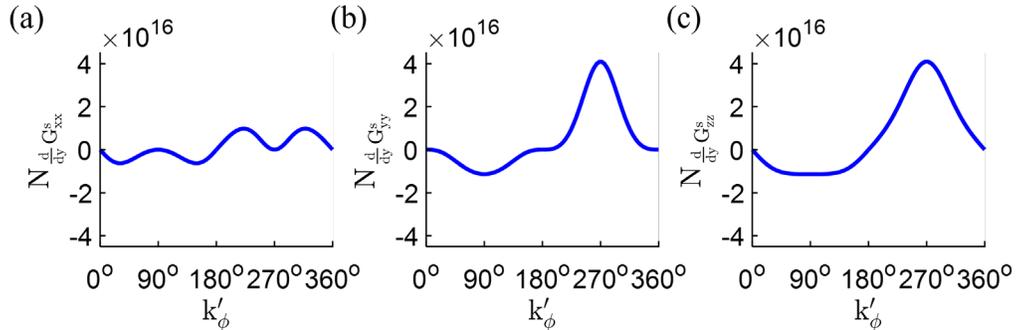

**Figure D1**: Residue of (a) $\frac{d}{dy}G_{xx}^s$, (b) $\frac{d}{dy}G_{yy}^s$, and (c) $\frac{d}{dy}G_{zz}^s$ at $\mathbf{k}_\rho$ versus $\mathbf{k}'_\phi$ calculated for a source point located over drift-biased nonreciprocal graphene. Results are computed using Eq. (D2) keeping $v_d = 0.5 v_F$. Other parameters are as in Figure 1.

It should be stressed that Eq. (D3) captures the fundamental physics of the problem: lateral recoil forces mostly depend on the dispersion relation of the modes supported by the platform. Additionally, as shown in Figure D1 for a specific example, the integrands of the functions shown in Eq. (D3) exhibit a smooth, non-singular behavior versus $k'_\phi$ and thus can easily be integrated using numerical routines [53]. A closer look into Eqs. (D2)-(D3) reveals that $\text{Re}\left\{\frac{d}{dy}G^s_{xx}\right\}$ is strictly zero along/against the bias direction, whereas the terms $\text{Re}\left\{\frac{d}{dy}G^s_{yy}\right\}$ and $\text{Re}\left\{\frac{d}{dy}G^s_{zz}\right\}$ are minimum/maximum along/against such direction. The former response is expected, because in $\text{Re}\left\{\frac{d}{dy}G^s_{xx}\right\}$ the bias is applied in the direction orthogonal to the *xx*-tensor component of the Green's functions and therefore it is exactly zero along the *y*-direction. The latter response appears because the momentum of the supported modes along/against the bias is minimum/maximum [37-40]. Additionally, Eq. (D3) shows a dependence $\propto k_\rho^6$ and therefore it is expected that modes with larger momentum will strongly dominate the platform overall response. Rooted on these arguments, we derive an approximate analytical expression for Eq. (D1). To this purpose, we consider first that the power scattered by the particle in the direction orthogonal to the bias is not dominant, i.e., $P^x_{\text{rad}} \leq P^y_{\text{rad}}$ and $P^x_{\text{rad}} \leq P^z_{\text{rad}}$. That condition is relatively general and appears in most practical scenarios. One exception occurs when a plane wave quasi-vertically illuminates the platform with a polarization aligned in the direction orthogonal to the bias ($\hat{x}$-axis in this case). Assuming that we are not in that specific situation, Eq. (D1) can be simplified to

$$F_y^{\text{nr-a}} \approx \frac{6\pi}{c_0 k_0^2}\left[P^y_{\text{rad}}\text{Re}\left\{\frac{d}{dy}G^s_{yy}\right\} + P^z_{\text{rad}}\text{Re}\left\{\frac{d}{dy}G^s_{zz}\right\}\right], \quad (D4)$$

where we have assumed that $P^y_{\text{rad}}\text{Re}\left\{\frac{d}{dy}G^s_{yy}\right\} \gg P^x_{\text{rad}}\text{Re}\left\{\frac{d}{dy}G^s_{xx}\right\}$ and $P^z_{\text{rad}}\text{Re}\left\{\frac{d}{dy}G^s_{zz}\right\} \gg P^x_{\text{rad}}\text{Re}\left\{\frac{d}{dy}G^s_{xx}\right\}$. To calculate the real part of the spatial derivatives of the Green's function tensor, we evaluate Eqs. (D3b)-(D3c) along and against the bias direction. This approximation assumes a delta function at these directions and is justified due to the $\propto k_\rho^6$ dependence of the force on the momentum of the supported states. Since the momentum of the states along and against the bias are maximum in their respective semi-plane of the momentum space, they are expected to dominate the response of the platform. Following this strategy, the spatial derivatives of the Green's functions required in Eq. (D4) can be computed analytically as

$$\text{Re}\left\{\frac{d}{dy}G_{yy}^s\right\} \approx \frac{1}{16\varepsilon_r k_0^2}\left[\left(k_y^-\right)^4 e^{-2z_0\sqrt{(k_y^-)^2-k_0^2}} - \left(k_y^+\right)^4 e^{-2z_0\sqrt{(k_y^+)^2-k_0^2}}\right], \quad \text{(D5a)}$$

$$\text{Re}\left\{\frac{d}{dy}G_{zz}^s\right\} \approx \frac{1}{16\varepsilon_r k_0^2}\left[\left(k_y^-\right)^4 e^{-2z_0\sqrt{(k_y^-)^2-k_0^2}} - \left(k_y^+\right)^4 e^{-2z_0\sqrt{(k_y^+)^2-k_0^2}}\right]. \quad \text{(D5b)}$$

Substituting the compact form expression of these derivatives into Eq. (D4), the nonreciprocity-induced recoil force acting along the external bias direction yields the analytical expression shown in Eq. (10).

**D2. Lateral recoil forces due to the broken symmetry of the system in polarization**

The other component of the nonreciprocity induced lateral recoil force, $F_x^{nr-p}$, acts along the orthogonal direction with respect to the applied momentum bias [see Eq. (7a)] and appears due to the broken symmetry in polarization of the supported nonreciprocal SPPs [31]. To calculate this component, one needs first to calculate the real part of the Green's functions spatial derivative at the particle position, i.e., $\text{Re}\left\{\frac{d}{dx}G_{xy}^s\right\}$. Assuming the nonretarded regime, and applying the approach described above, the integral along $k_\rho'$ of this derivative at the particle position can be computed as

$$N_{\frac{d}{dx}G^S_{xy}}(k'_\phi) \approx -i\frac{k_\rho^4 \cos k'_\phi \sin(2k'_\phi)}{32\pi\varepsilon_r k_0^2}e^{-2z_0\sqrt{k_\rho^2-k_0^2}}, \tag{D6}$$

where again the $k_\rho - k'_\phi$ relationship is determined by the dispersion relation of the system. Then, the spatial derivative of the Green's functions is computed at the particle position by performing the polar integral along $k'_\phi$:

$$\text{Re}\left\{\frac{d}{dx}G^S_{xy}\right\} = \int_0^{2\pi} \frac{k_\rho^4 \cos k'_\phi \sin(2k'_\phi)}{32\varepsilon_r k_0^2}e^{-2z_0\sqrt{k_\rho^2-k_0^2}}dk'_\phi. \tag{D7}$$

The integrand of this function exhibits a smooth behavior versus $k'_\phi$ and can be integrated using standard numerical routines [53]. Importantly, Eq. (D7) reveals a dependence $\propto k_\rho^4$ with the momentum of the supported states, two orders of magnitude smaller than the one found in Eq. (D3) for recoil forces arising from the broken symmetry of the system in amplitude. Combining Eqs. (D7) and (7a) permits to express this recoil force component as in Eq. (9a).

**D3. Lateral recoil forces due to spin-orbit effects**

The lateral recoil forces $\mathbf{F}^S$ in Eq. (7c) appear due to the spin-orbit effect of light. To compute these force components, one needs to compute the spatial derivatives of the Green's functions at the particle position, i.e., $\text{Im}\left\{\frac{d}{dx}G^S_{xz}\right\}$ and $\text{Im}\left\{\frac{d}{dy}G^S_{yz}\right\}$. Following a similar approach as described above, and assuming that the platform is operated in the nonretarded regime, the residues of these derivatives for a specific azimuthal direction $k'_\phi$ can be computed as

$$N_{\frac{d}{dx}G^S_{xz}}(k'_\phi) \approx -i\frac{1}{8\varepsilon_r \pi^2 k_0^2}k_x^4 e^{-2z_0\sqrt{k_x^2-k_0^2}}, \tag{D8a}$$

$$N_{\frac{d}{dy}G^S_{yz}}(k'_\phi) \approx -i\frac{1}{8\varepsilon_r \pi^2 k_0^2}k_y^4 e^{-2z_0\sqrt{k_y^2-k_0^2}}. \tag{D8b}$$

Next, the spatial derivatives of the Green's functions required is computed at the particle position by performing the polar integrals along $k'_\phi$:

$$\text{Im}\left\{\frac{d}{dx} G^s_{xz}\right\} = \int_0^{2\pi} \frac{1}{8\varepsilon_r \pi k_0^2} k_x^4 e^{-2z_0\sqrt{k_x^2-k_0^2}} dk'_\phi, \qquad (D9a)$$

$$\text{Im}\left\{\frac{d}{dy} G^s_{yz}\right\} = \int_0^{2\pi} \frac{1}{8\varepsilon_r \pi k_0^2} k_y^4 e^{-2z_0\sqrt{k_y^2-k_0^2}} dk'_\phi. \qquad (D9b)$$

These equations can be easily integrated over $k'_\phi$, yielding

$$\text{Im}\left\{\frac{d}{dx} G^s_{xz}\right\} \approx \frac{1}{16\varepsilon_r k_0^2}\left[(k_x^-)^4 e^{-2z_0\sqrt{(k_x^-)^2-k_0^2}} + (k_x^+)^4 e^{-2z_0\sqrt{(k_x^+)^2-k_0^2}}\right], \qquad (D10a)$$

$$\text{Im}\left\{\frac{d}{dy} G^s_{yz}\right\} \approx \frac{1}{16\varepsilon_r k_0^2}\left[(k_y^-)^4 e^{-2z_0\sqrt{(k_y^-)^2-k_0^2}} + (k_y^+)^4 e^{-2z_0\sqrt{(k_y^+)^2-k_0^2}}\right], \qquad (D10b)$$

where $k_x^-$ and $k_x^+$ are the plasmon wavenumber in the negative and positive $k_x$-half spaces, respectively, in the direction orthogonal to the external bias. Combining Eqs. (D10) and (7c) permits to analytically express these recoil force components as in Eq. (13).

It is important to stress that the strength of spin-orbit recoil forces is usually weak when the particle is illuminated by a linearly polarized light [10]. This is because the particle acquires a weak polarization spin due to the lack of spin of the incident light.